%% file: main.tex
\documentclass[journal,twoside,web]{style/ieeecolor}

\usepackage{etoolbox}
\makeatletter
\@ifundefined{color@begingroup}%
  {\let\color@begingroup\relax
   \let\color@endgroup\relax}{}%
\def\fix@ieeecolor@hbox#1{%
  \hbox{\color@begingroup#1\color@endgroup}}
\patchcmd\@makecaption{\hbox}{\fix@ieeecolor@hbox}{}{\FAILED}
\patchcmd\@makecaption{\hbox}{\fix@ieeecolor@hbox}{}{\FAILED}

\usepackage{style/tbme}

\usepackage{cite}
\usepackage{amsmath, amssymb, amsfonts}
\usepackage{graphicx}
\usepackage{textcomp}
\usepackage{afterpage}

\usepackage{algorithm}
\usepackage{algpseudocode}
\usepackage{multirow}
\usepackage{tabularx}
\usepackage{booktabs}
\usepackage{rotating}
\usepackage{siunitx}
\usepackage{subcaption}
\usepackage{graphicx}
\usepackage{makecell}
\usepackage{url}    %
\usepackage{xcolor} %
\usepackage{ifthen} %
\usepackage{ulem}
\usepackage{mathtools}
\mathtoolsset{showonlyrefs}

\newcounter{highlight}
\input{math.tex}

\input{acronyms.tex}

\def\BibTeX{{\rm B\kern-.05em{\sc i\kern-.025em b}\kern-.08em
    T\kern-.1667em\lower.7ex\hbox{E}\kern-.125emX}}
\usepackage{url}
\makeatletter
\let\NAT@parse\undefined
\makeatother
\usepackage[hidelinks]{hyperref}

\begin{document}

    \onecolumn
    \thispagestyle{empty}
    \begin{center}
        {\Large Physics-Informed Deep Learning for Shear Wave Speed Estimation in MR Elastography\par}
        \vspace{0.6em}
        {\large S. Martin, J. Schattenfroh, P. Schuenke, F. F. Zimmermann, I. Sack, C. Kolbitsch, A. Kofler\par}
    \end{center}
    \vspace{1.5em}
    \noindent\copyright~2026 The Authors. This work is licensed under a Creative Commons Attribution 4.0 License. For more information, see \url{https://creativecommons.org/licenses/by/4.0/}.

    \vspace{0.8em}
    \noindent This is the authors' accepted version of an article published in IEEE Transactions on Biomedical Engineering. Changes were, or will be, made to this version by the publisher prior to final publication.

    \vspace{0.8em}
    \noindent DOI: \href{https://doi.org/10.1109/TBME.2026.3666306}{10.1109/TBME.2026.3666306}\\
    IEEE Xplore: \url{https://ieeexplore.ieee.org/document/11407998}

    \vspace{1.5em}
    \noindent Please cite this paper as:\\
    S. Martin, J. Schattenfroh, P. Schuenke, F. F. Zimmermann, I. Sack, C. Kolbitsch, and A. Kofler, ``Physics-Informed Deep Learning for Shear Wave Speed Estimation in MR Elastography,'' in \textit{IEEE Transactions on Biomedical Engineering}, 2026, doi: \href{https://doi.org/10.1109/TBME.2026.3666306}{10.1109/TBME.2026.3666306}.

    \vspace{1.5em}
    \noindent Code: \url{https://github.com/Stef-Martin/E2E-MRE}

    \setcounter{page}{0}
    \twocolumn

    \title{Physics-Informed Deep Learning for Shear Wave Speed Estimation in MR Elastography}

    \author{
        Stefan Martin, Jakob Schattenfroh, Patrick Schuenke, Felix Frederik Zimmermann,\\ Ingolf Sack, Christoph Kolbitsch, Andreas Kofler
                \thanks{Date of submission: Jan. 09, 2026. This paper was supported by the German Research Foundation (GRK2260–BIOQIC, CRC1340-Matrix in Vision). 
        This work was in part supported by the Federal Ministry for Economic Affairs and Energy (BMWi) in the frame of the QI-Digital initiative through the Metrology for Artificial Intelligence for Medicine (M4AIM) project, in part by project 22HLT02 A4IM which has received funding from the European Partnership on Metrology, co-financed from the European Union’s Horizon Europe Research and Innovation Programme and by the Participating States. The funding is gratefully acknowledged.}
        \thanks{
            S. Martin, P. Schuenke, F.F. Zimmermann, C. Kolbitsch and A. Kofler are with Physikalisch-Technische Bundesanstalt (PTB), 
            Braunschweig and Berlin, Germany (e-mails: stefan.martin@ptb.de, patrick.schuenke@ptb.de, felix.zimmermann@ptb.de, christoph.kolbitsch@ptb.de, andreas.kofler@ptb.de).
            }
        \thanks{
            J. Schattenfroh and I. Sack are with Charité - Universit\"{a}tsmedizin Berlin, Germany
            (e-mails: jakob.schattenfroh@charite.de, ingolf.sack@charite.de)
        }
    }

    \maketitle

    \begin{abstract}

        \textit{Objective:} Magnetic Resonance Elastography (MRE) is a non-invasive imaging technique for mapping biomechanical properties of in vivo tissue, including shear wave speed (SWS), but involves intrinsically slow data acquisition and an ill-posed wave inversion. Instead of relying on handcrafted image priors, we propose a data-driven approach jointly combining image reconstruction and MRE inversion for robust SWS estimation from undersampled \textit{k}-space data. \textit{Methods:} Our physics-informed reconstruction framework comprises two blocks: a model-based neural network (NN)-regularized reconstruction module and a phase-gradient inversion (\textit{k}-MDEV) calculating SWS from the reconstructed images. Concatenating both blocks yields an end-to-end trainable method to estimate SWS directly from measured \textit{k}-space data. We evaluated the method on retrospectively highly undersampled brain MRE data and compared it to a total variation (TV) minimization-based approach. We assessed the impact of end-to-end training (qualitative images \textit{and} SWS maps as targets) versus pre-training (qualitative images as targets) and applied the method also to in vivo data. \textit{Results}: Our approach significantly reduces NRMSE by 30\% compared to TV. End-to-end training improves SWS estimation over separate image reconstruction and SWS calculation. \textit{Conclusion:} Accurate SWS quantification is possible at acceleration factors up to 19. Our method significantly outperforms TV, highlighting the need for data-driven regularization in this challenging MR problem. Further, our approach successfully generalizes to in vivo data. \textit{Significance:} We present the first end-to-end trainable MRE reconstruction method for estimating SWS maps directly from \textit{k}-space. NN-based reconstruction can enable rapid stiffness mapping for dynamic studies, functional imaging, and real-time clinical feedback.

    \end{abstract}
    
    \begin{IEEEkeywords}
        Magnetic Resonance Elastography, End-to-End Learning, Physics-Informed Deep Learning
    \end{IEEEkeywords}
    
    \input{sections/1_introduction.tex}
    \input{sections/2_problem.tex}
    \input{sections/3_methods.tex}
    \input{sections/4_experiments.tex}
    \input{sections/5_results.tex}
    \input{sections/6_discussion.tex}
    \input{sections/7_conclusion}

    \normalem
    \bibliography{bibliography.bib}

\end{document}

%% file: math.tex
\newcommand{\AHop}{\mathbf{A}^{\scriptstyle \boldsymbol{\mathsf{H}}}}

\newcommand{\herm}{{\scriptstyle \boldsymbol{\mathsf{H}}}}
\newcommand{\trans}{{\scriptstyle \boldsymbol{\mathsf{T}}}}

\usepackage{empheq}

\definecolor{r1}{RGB}{0,0,255} %
\definecolor{r2}{RGB}{255,0,0} %
\newcommand{\revision}[2]{%
  \ifnum\value{highlight}=1%
    \ifnumequal{#1}{1}{{\color{r1}{#2}}}{}%
    \ifnumequal{#1}{2}{{\color{r2}{#2}}}{}%
  \else%
    #2%
  \fi%
}

\newcommand{\delete}[2]{%
  \ifnum\value{highlight}=1%
    \ifnumequal{#1}{1}{{\color{r1}{\sout{#2}}}}{}%
    \ifnumequal{#1}{2}{{\color{r2}{\sout{#2}}}}{}%
  \fi%
}

%% file: acronyms.tex
\usepackage[abbreviations]{glossaries-extra}
\glssetcategoryattribute{acronym}{indexonlyfirst}{true}

\newabbreviation{mri}{MRI}{magnetic resonance imaging}
\newabbreviation{mr}{MR}{magnetic resonance}
\newabbreviation{fov}{FoV}{field of view}
\newabbreviation{gpu}{GPU}{graphical processing unit}

\newabbreviation{nn}{NN}{neural network}
\newabbreviation{cnn}{CNN}{convolutional neural network}
\newabbreviation{fft}{FFT}{fast Fourier transform}
\newabbreviation{cg}{CG}{conjugate gradient}
\newabbreviation{sh}{SH}{spherical harmonic}
\newabbreviation{shn}{SH-Net}{spherical-harmonic network}
\newabbreviation{orn}{ORN}{off-resonance-network or OffResNet}
\newabbreviation{pdhg}{PDHG}{primal dual hybrid gradient}

\newabbreviation{mse}{MSE}{mean squared error}
\newabbreviation{rmse}{RMSE}{root mean squared error}
\newabbreviation{nrmse}{NRMSE}{normalized root mean squared error}
\newabbreviation{psnr}{PSNR}{peak signal-to-noise ratio}
\newabbreviation{snr}{SNR}{signal-to-noise ratio}
\newabbreviation{mae}{MAE}{mean absolute error}
\newabbreviation{ssim}{SSIM}{structural similarity index}
\newabbreviation{roi}{RoI}{region of interest}
\newabbreviation{iqr}{IQR}{interquartile range}

%% file: sections/1_introduction.tex
\newpage
\section{Introduction} \label{sec:introduction}
\IEEEPARstart{M}{agnetic} Resonance Imaging (MRI) is a versatile and powerful imaging technique that allows the acquisition of high-resolution qualitative images of the internal anatomy of the human body. In quantitative MRI (qMRI), various quantitative parameters with specific biophysical correspondences can then be estimated from the image data. MR Elastography (MRE) \cite{muthupillai1995magnetic,manduca2001magnetic,hirsch2017magnetic} is a qMRI technique that provides detailed maps of biomechanical properties, enabling a non-invasive assessment of tissue stiffness. Externally induced mechanical waves result in tissue displacements, which are encoded as proportional phase shifts in the MR signal by motion encoding gradients (MEG) that are incorporated into specially designed MRI sequences. These phase shifts are tracked over time, enabling the calculation of shear wave speed (SWS)\cite{sack2023magnetic}, which generally involves three steps: acquisition of $k$-space data, reconstruction of complex-valued MR images from the data, and estimation of quantitative SWS maps from these images with an inversion method. 

The brain is a fascinating target for MRE, particularly for detecting and monitoring neurological disorders that are associated with localized changes in tissue stiffness, such as Alzheimer’s disease\cite{murphy2016regional}. Conventional MRE techniques typically require multiple time-resolved acquisitions, resulting in extended acquisition protocols with prolonged scan times and vibration exposure. These factors can increase patient discomfort, reduce clinical efficiency, and limit temporal resolution, making it difficult to investigate dynamic physiological processes that require fast, continuous measurements. Real-time MRE addresses these limitations by enabling high frame-rate stiffness mapping\cite{schrank2020real}. This approach improves patient comfort, reduces motion-related artifacts, and enables the monitoring of dynamic changes in brain tissue properties -- such as those occurring during the Valsalva maneuver\cite{herthum2021real} or cardiac-cycle-related brain pulsations, as demonstrated in ultrasound elastography\cite{meyer2023stiffness}. Recent studies have shown the feasibility and clinical value of rapid MRE in various applications, including subsecond mapping of liver stiffness\cite{anders2024rapid}, evaluation of the response in skeletal muscle to loading at frame rates of $\sim6\,\text{Hz}$\cite{schrank2020real}, and functional brain MRE\cite{patz2019imaging}.

Real-time stiffness mapping can be achieved by reducing scan time through undersampled data acquisition. In this work, we utilize a motion-robust single-shot spiral readout for MRE data acquisition, enabling the capture of all required data in a single shot per timestep and MEG. However, data undersampling introduces challenges into MRI reconstruction, yielding an ill-posed inverse problem necessitating the use of regularization techniques to stabilize the reconstruction. Importantly, the subsequent generation of MRE maps from wave images is itself inherently ill-posed due to noise and unknown boundary conditions.

Undersampled MRE images have been reconstructed using established non-learning approaches, which are typically applied prior to elastography inversion. These include parallel imaging and compressed sensing with sparsity priors, such as MRE-specific phase or displacement regularization \cite{mohammed_phaseregularized_2023}, as well as Bayesian models that impose inter–offset constraints \cite{ebersole_bayesian_2018}. Further, machine learning (ML) has significantly influenced and improved MR image reconstruction \cite{wang2018image} in recent years and currently defines the state-of-the-art both in MRI and qMRI reconstruction, see e.g.\ \cite{kofler2024machinelearningquantitativemr} for an extensive review. Despite these advances, several challenges persist. Frame rates in current rapid MRE techniques are still limited, and developments in neural network (NN)-based reconstruction methods and real-time MRE have largely progressed as separate fields. This work integrates both approaches to enhance the stiffness mapping frame rate while maintaining the accuracy of the reconstructed MR images and SWS maps. So far, ML in MRE has been applied to segmentation tasks \cite{aldoj2022fully}, phase unwrapping optimization \cite{ma2023mr}, and learning the inversion method for the calculation of stiffness maps\cite{bustinneural,scott2020artificial,murphy2018artificial,yin2023swenet}.

In this work, we employ a NN-based regularization in a physics-informed image reconstruction framework \cite{schlemper2017deep,aggarwal2018modl} that can be trained end-to-end \cite{kofler2021end,zimmermann2024pinqi} -- from $k$-space data to stiffness map. By treating qualitative image reconstruction and SWS quantification as joint tasks, we enable the NN to learn a task-specific regularization tailored to the quantification of SWS by the inversion method $k$-MDEV \cite{tzschatzsch2016tomoelastography}.

%% file: sections/2_problem.tex
\section{Problem Formulation} \label{sec:problem}
The process of data acquisition in MRI is typically modeled as
\begin{equation}\label{eq:forward_model}
    \mathbf{y}=\mathbf{A}\mathbf{x}_{\mathrm{true}}+\mathbf{e}.
\end{equation}
Here, $\mathbf{y}$ represents the measured $k$-space data, $\mathbf{A}$ the MRI operator, $\mathbf{x}_\text{true}$ the complex-valued MR image that has to be reconstructed, and $\mathbf{e}$ denotes Gaussian noise, which accounts for measurement errors.
In qMRI, the considered forward problem is extended by performing several different measurements of the same underlying quantity, thereby collecting complementary data by, for example, applying different preparation pulses or by varying the MR sequence parameters. Formally, this can be described by a (typically non-linear) model $q_s$, which models all relevant scanner and MR sequence parameters defined by $s$ (see e.g. \cite{tamir2020computational}). 
It  relates the quantities of interest to the prepared image, i.e.
\begin{equation}\label{eq:signal_model}
    \mathbf{x}_s:= q_s(\mathbf{p}_{\mathrm{true}}) + \boldsymbol{\eta}_s,
\end{equation}
where $\mathbf{p}_{\mathrm{true}}$ is the reference parameter associated with the reference image $\mathbf{x}_{\mathrm{true}}$ and $\boldsymbol{\eta}_s$ accounts for non-trivial modeling errors inherently present in $q_s$. The corresponding $k$-space data are acquired according to \eqref{eq:forward_model}.
In MRE, specifically, the model $q_s$ describes the wave propagation as reflected in tissue displacement. The entire acquisition can thus be expressed as
\begin{equation}\label{eq:acq_mre}
    \mathbf{y}=\mathbf{A}q_{\mathcal{S}}(\mathbf{p})+\mathbf{e},
\end{equation}
where $q_{\mathcal{S}}:\mathbf{p} \mapsto [q_{s_1}(\mathbf{p}), q_{s_2}(\mathbf{p}),\ldots, q_{s_{ |\mathcal{S}|}}(\mathbf{p})]^\trans$ and each  $q_{s_i}(\mathbf{p})$ represents the $i$-th wave field generated by the forward model with set-up parameters defined by $s_i \in \mathcal{S}$ (more details on that later in Subsection \ref{subsec:image_reconstruction}). Note that the employed forward models $\mathbf{A}$ and $q_{\mathcal{S}}$ are assumed to be sufficiently accurate relative to the true models, such that the modeling error term $\mathbf{A} \boldsymbol{\eta}_s$ is considered negligible compared to the sensor noise $\mathbf{e}$ and, therefore, omitted from the acquisition model of \eqref{eq:acq_mre}. External mechanical vibrations are applied, generating shear waves propagating through the tissue. The resulting displacement depends on the underlying viscoelastic properties, which are characterized by the complex shear modulus $G^*$. The motion is captured by MEGs and encoded into the phase of the MR signal, which is read out during MR acquisition. The forward model $q_s$ in \eqref{eq:signal_model} maps the viscoelastic parameters to the measured MR phase, reflecting shear wave propagation. This wave is governed by the Navier equation, which describes the propagation of waves in a medium.
With several simplifying key assumptions -- time-harmonic motion, incompressibility (neglecting longitudinal wave components), and local homogeneity of the viscoelastic parameters -- we can simplify the Navier equation to the Helmholtz equation:
\begin{equation}
    \Delta \mathbf{u} +\frac{\omega^2}{\mathbf{c}[l]^2} \mathbf{u}=0,
    \label{eq:wave_eq}
\end{equation}
where $\mathbf{u}$ is the shear wave field (obtained from processed MR phase data: applying filters or the curl operator to suppress compressional components), $\omega=2\pi f$ is the angular frequency of the vibration, and $\mathbf{c}[l]$ is the SWS at voxel $l$. The homogeneity assumption allows spatial derivatives of material parameters to be neglected and enables this formulation.
The inverse problem, which is the one of interest in our application, aims to estimate the viscoelastic parameters from the shear wave field $\mathbf{u}$ and is commonly approached in three ways\cite{sack2023magnetic,mcgarry2011comparison}. One approach solves a weak formulation via finite elements (FEM); another directly inverts the Helmholtz equation and typically invokes local homogeneity; and a third relies on phase‑gradient or wavenumber‑based methods. Many FEM and NN approaches do not require the homogeneity assumption and explicitly accommodate heterogeneous media by solving full/weak forms of the governing equations or learning mappings from displacement fields to spatially varying properties. In this work, we use a phase-gradient inversion such as $k$-MDEV \cite{tzschatzsch2016tomoelastography}. It assumes locally planar waves and homogeneity and estimates the clinically relevant SWS from the displacement field encoded in the MR image phase. The SWS is given as $\mathbf{c}[l]=\frac{1}{\operatorname{Re\big(\sqrt{\rho /G^\ast[l]}\big)}}$, with the bulk density $\rho$ and the complex shear modulus $G^\ast[l]$ at voxel $l$. Consequently, for our problem, the vector of the parameters sought in \eqref{eq:acq_mre} is given by $\mathbf{p} = \mathbf{c}$.

%% file: sections/3_methods.tex
\section{Methods} \label{sec:methods}
Since problem \eqref{eq:acq_mre} is typically non-convex due to the non-linear model $q_{\mathcal{S}}$ and large-scale due to the acquisition of complementary measurements that allow for the quantification of $\mathbf{p}$, different solution strategies exist to tackle the problem, see \cite{kofler2024machinelearningquantitativemr}. 
In this work, we consider the widely used two-step approach for solving qMRI problems, i.e., decoupling the image reconstruction from the parameter quantification problem by
\begin{empheq}[left=\empheqlbrace]{alignat=1} 
        &\underset{\mathbf{p}}{\min} \,\frac{1}{2} \| q_{\mathcal{S}}(\mathbf{p}) - \mathbf{x}_{\mathrm{rec}}\|_2^2 \label{eq:param_estimation}\\
        &\mathbf{x}_{\mathrm{rec}}:=\underset{\mathbf{x}}{\mathrm{argmin}} \frac{1}{2}\| \mathbf{A}\mathbf{x} - \mathbf{y}\|_2^2 + \mathcal{R}(\mathbf{x}) \label{eq:image_recon}.
\end{empheq}
Thereby, \eqref{eq:image_recon} corresponds to reconstructing an estimate of the image $\mathbf{x}_{\mathrm{rec}}\approx\mathbf{x}_{\mathrm{true}}$ in the ill-posed inverse problem in \eqref{eq:forward_model}, and \eqref{eq:param_estimation} addresses the estimation of the relevant quantitative parameters $\mathbf{p}_{\mathrm{true}}$ from the reconstructed image data as in \eqref{eq:signal_model}. Based on the assumption that the only measurement error is sensor noise $\mathbf{e}$, the data-fidelity term in \eqref{eq:image_recon} is formulated using an $\ell_2$-norm, which corresponds to the log-likelihood function under an additive Gaussian noise model.
Note that the noise component in \eqref{eq:signal_model} is contained in $\mathbf{x}_{\mathrm{rec}}$ in \eqref{eq:param_estimation} and is inherently linked to the choice of the regularization method $\mathcal{R}$ in \eqref{eq:image_recon} and its capability to approximate the unknown target image. When solving \eqref{eq:image_recon} and \eqref{eq:param_estimation} in a sequential fashion, once the image data $\mathbf{x}_{\mathrm{rec}}$ in \eqref{eq:image_recon} are fixed after the reconstruction step, residual errors directly impact the quantification process in \eqref{eq:param_estimation}. As such, it is crucial to choose an expressive regularization method $\mathcal{R}$ that can represent complex priors or constraints on the solution.

Here, we propose to construct an end-to-end trainable NN based on algorithm unrolling \cite{monga2021algorithm} that comprises two main blocks. The first block consists of a model-based deep learning approach \cite{aggarwal2018modl} that constructs a reconstruction network alternating between CNN application and data consistency (DC) layers. 
The second block corresponds to the $k$-MDEV inversion method\cite{tzschatzsch2016tomoelastography} for computing SWS maps from the reconstructed image data. Notably, both blocks can be trained end-to-end, i.e.\, from $k$-space data to SWS maps. As a result, the regularizer $\mathcal{R}$ in \eqref{eq:image_recon} is optimized for the downstream task of SWS quantification -- the final quantity of interest in MRE -- rather than for reconstructing precise qualitative images, which serve only as an intermediate step in the parameter estimation.

\subsection{Image Reconstruction}\label{subsec:image_reconstruction}
First, we expand upon the forward problem introduced in \eqref{eq:forward_model} to precisely formulate our method for solving problem \eqref{eq:image_recon}. We consider the general setup of multi-coil MRI, where the $k$-space is acquired using multiple receiver coils to be able to accelerate the acquisition by data undersampling while, at the same time, maintaining comparable image quality when reconstructing the images.

In the specific case of MRE, $N_{\mathrm{time}}$ vibration phase offsets are typically used to track wave propagation over time. Additionally, $N_{\mathrm{MEG}}$ MEG directions and $N_{\mathrm{freq}}$ different frequencies can be acquired per slice and per subject, allowing for a more precise SWS computation, as demonstrated in \cite{hirsch2017magnetic}. The image data from these various MEG directions, frequencies, and vibration phase offsets can then be used to compute a single SWS map.
For a fixed MEG $m$, frequency $f$ and time point $t$, we consider the linear operator $\mathbf{A}_I$, which maps a complex-valued 2D image $\mathbf{x}\in \mathbb{C}^N$ to its corresponding multi-coil $k$-space data by
\begin{equation}\label{eq:forward_operator_single_time}
\begin{split}
    \mathbf{A}_I: \,\mathbb{C}^N &\rightarrow \mathbb{C}^M, \\
    \mathbf{x} &\mapsto  \mathbf{A}_I \mathbf{x} :=(\mathbf{I}_{N_{\mathrm{coils}}} \otimes \mathbf{E}_I)\mathbf{C} \,\mathbf{x},    
\end{split}
\end{equation}
where $\mathbf{C}=[\mathbf{C}_1,\ldots, \mathbf{C}_{N_{\mathrm{coils}}}]^\trans$ with $\mathbf{C}_j:=\mathrm{diag}(\mathbf{c}_j)$ with $\mathbf{c}_j \in \mathbb{C}^N$ denotes the $j$-th coil sensitivity map (CSM), $\mathbf{E}_I$ is a 2D Fourier operator that maps a 2D image to its $k$-space coefficients indexed by the set $I\subset J=\{1,\ldots,N_{k{\text{-space}}}\}$, $M:=N_{\mathrm{coils}} \cdot |I|$ and $\mathbf{I}_{N_{\mathrm{coils}}}$ represents a $N_{\mathrm{coils}}\times N_{\mathrm{coils}}$-identity operator. Note that in this work, $\mathbf{E}_I$ corresponds to a non-uniform fast Fourier operator %
that samples $k$-space data along spiral trajectories.
In our approach, to be able to exploit the correlation between the images acquired for different time points $t_i$, the precise form of the operator in \eqref{eq:forward_model} is 
\begin{equation}\label{eq:forward_operator_all_times}
                \mathbf{A}\mathbf{x}:= \bigg(\bigoplus_{i=1}^{N_{\mathrm{time}}} \mathbf{A}_{I_i}\bigg) \mathbf{x} :=[\mathbf{A}_{I_1}\mathbf{x}_1, \ldots, \mathbf{A}_{I_{N_{\mathrm{time}}}} \mathbf{x}_{N_{\mathrm{time}}}]^\trans,
\end{equation}
with each $\mathbf{x}_i \in \mathbb{C}^{N}$ and $\mathbf{A}_{I_i}$ given in the form of \eqref{eq:forward_operator_single_time}, meaning that we tackle the reconstruction of the data simultaneously for all time points, i.e. $q_{\mathcal{S}}=q_{\mathcal{T}}$ with $\mathcal{T}:=\{t_1, \ldots, t_{N_{\mathrm{time}}}\}$, but separately for the different frequencies $f \in \{f_1, \ldots, f_{N_{\mathrm{freq}}}\}$ and MEGs $m \in \{m_1, \ldots, m_{N_{\mathrm{MEG}}}\}$. Crucially, the sets of indices $I_1,\ldots, I_{N_{\mathrm{time}}}$ vary for different acquisition time points $t_1, \ldots, t_{N_{\mathrm{time}}}$, thus yielding complementary information in the $k$-space data, which can be used for regularization. Consequently, the problem in \eqref{eq:forward_model} is rather large, involving $k$-space data that are acquired with multiple receiver coils and for different time steps, requiring a computationally efficient reconstruction algorithm.

We approach the reconstruction of the undersampled MRE images in \eqref{eq:image_recon} by a model-aware reconstruction network that follows MoDL \cite{aggarwal2018modl}, i.e.\, given an initial reconstruction $\mathbf{x}_0$, we construct an alternating sequence of iterates 
\begin{empheq}[left={\hspace{-0.5em}\empheqlbrace}]{alignat=1} 
        &\mathbf{z}_k := \mathrm{u}_{\theta}(\mathbf{x}_k) \label{eq:net_application}\\
        &\mathbf{x}_{k+1}:=\underset{\mathbf{x}}{\mathrm{argmin}} \frac{1}{2}\| \mathbf{W}^{1/2}(\mathbf{A}\mathbf{x} - \mathbf{y})\|_2^2 + \frac{\lambda_k}{2}\| \mathbf{x} - \mathbf{z}_k\|_2^2 \label{eq:dc_update},
\end{empheq}
of length $T$, where $u_{\theta}$ denotes a learnable network and $\lambda_k>0$ are learnable scalar parameters. The term $\frac{1}{2}\|\mathbf{W}^{1/2}(\mathbf{A}\mathbf{x}-\mathbf{y})\|_2^2$ enforces DC between estimated image and measured data $\mathbf{y}$ in raw data space through a weighted $\ell_2$-norm involving a positive definite density compensation (DCF) operator $\mathbf{W}^{1/2}$ that corrects for non-uniform sampling via Voronoi tessellation. The Tikhonov regularization term $\frac{\lambda_k}{2}\| \mathbf{x} - \mathbf{z}_k\|_2^2$ enforces proximity of the current iterate to an NN-based image prior that is obtained as the output of an expressive CNN-block, reducing noise and artifacts. 
Since problem \eqref{eq:forward_model} is in general overdetermined due to multiple receiver coils, we can initialize the reconstruction network with an approximate pseudo-inverse solution obtained by approximately solving the normal equations, i.e.\ $\mathbf{x}_0:=\mathbf{x}^\dagger$
with $\mathbf{x}^\dagger:=(\AHop \mathbf{W} \mathbf{A})^{-1}\AHop \mathbf{W} \mathbf{y}$. Computing $\mathbf{x}_{k+1}$ in \eqref{eq:dc_update} corresponds to solving a linear system  $\mathbf{H}_k\mathbf{x}=\mathbf{b}_k$ of the form
\begin{equation}\label{eq:linear_system}
(\AHop \mathbf{W}\mathbf{A}+\lambda_k\mathbf{I})\mathbf{x}=\AHop \mathbf{W} \mathbf{y}+\lambda_k\mathbf{z}_k,
\end{equation}
which we tackle by means of a conjugate gradient (CG) method \cite{pruessmann2001advances}. We note that the number of iterations $T$ that are used to unroll the network defined by \eqref{eq:net_application} and \eqref{eq:dc_update} is a hyperparameter that impacts both the reconstruction quality as well as the GPU memory footprint and the training times of the considered method. In \cite{aggarwal2018modl}, it was shown that a larger $T$ tends to lower the achievable reconstruction error; however, it also increases training times and hardware requirements. In our case, we found $T=6$ to provide a good balance between reconstruction performance, GPU memory usage, and training time. Further, it is worth noting that setting $T=1$ corresponds to a conceptually different approach, namely the one presented in \cite{kofler2020neural}, that -- although exhibiting some theoretically favorable properties (convexity of the thereby considered reconstruction problem \eqref{eq:image_recon}) -- empirically falls short compared to unrolling $T>1$ iterations, especially for more complex problems with high acceleration factor as the one considered in this work. 

\begin{figure*}
    \centering
    \includegraphics[width=\linewidth]{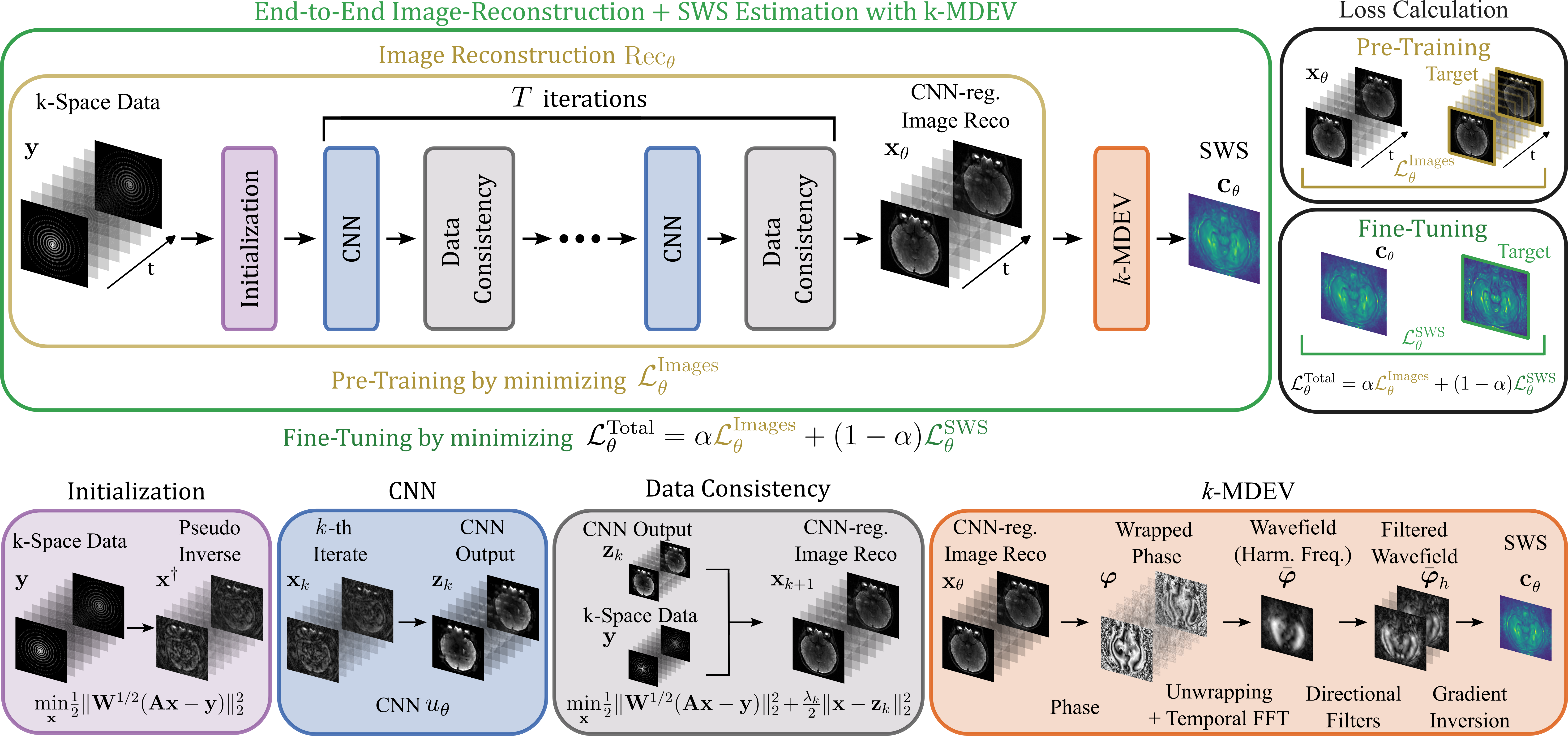}
    \caption{\textit{Yellow block}: Image reconstruction block $\mathrm{Rec}_{\theta}$: Starting from undersampled spiral $k$-space data, an initial MR image estimate is computed by approximating the pseudo-inverse solution (\textit{purple block}). \textit{Grey block}: An image is fed into a CNN, whose output is refined via a DC step according to \eqref{eq:dc_update}. CNN and DC steps are repeated $T$ times. Pre-training minimizes the loss between reconstructed and target images.
    \textit{Orange block}: $k$-MDEV: The phase is extracted, unwrapped, and temporally Fourier transformed. The harmonic component of the wave field is filtered and inverted to yield SWS maps, as described in \eqref{eq:kmdev_inversion}.
    \textit{Green block}: The entire reconstruction network, comprising the image reconstruction and the $k$-MDEV module, is fine-tuned by minimizing a combined loss that uses reference images and reference SWS maps.}
    \label{fig:Reco}
\end{figure*}

In our specific problem, we can leverage the temporal correlation in the phase of the complex MR images generated by the mechanical wave traveling through the tissue. Therefore, we use the network presented in \cite{kofler2019spatio}, which exploits the underlying spatio-temporal correlation in the data by extracting $xt$- and $yt$-spatio-temporal slices, applying a 2D U-Net \cite{ronneberger2015u} to reduce noise and artifacts, and reassembling the slices to the image volume. We refer to this network as XTYT. The framework of this physics-informed reconstruction network -- which incorporates the physical forward model $\mathbf{A}$ and the acquired $k$-space data $\mathbf{y}$ through DC blocks -- is denoted as $\mathrm{Rec}_{\theta}$ and is illustrated in Fig. \ref{fig:Reco}.

\subsection{\textit{k}-MDEV for SWS Estimation} \label{subsec:kmdev}
In the following, we briefly describe $k$-MDEV \cite{tzschatzsch2016tomoelastography}, which is a method that estimates SWS maps from the reconstructed image data, recovering all necessary wave information from the wave vector, and extending classical phase-gradient approaches\cite{manduca2001magnetic} by accounting for wave superposition and using directional filtering\cite{manduca2003spatio}. The method transitions from a continuous wave description to a discrete image-domain representation compatible with the nature of MR imaging. For details, see \cite{hirsch2017magnetic, tzschatzsch2016tomoelastography}.

\subsubsection{Phase}
Let $\mathbf{x} = |\mathbf{x}| \mathrm{e}^{\mathrm{i}\boldsymbol{\varphi}} \in \mathbb{C}^{N_{\mathrm{time}} \cdot N}$
denote the polar representation of the complex-valued MR image reconstructed as explained in \ref{subsec:image_reconstruction}. Mechanical vibrations induce shear waves, and MEGs encode this periodic motion proportionally into the image phase $\boldsymbol{\varphi}$, which approximates the displacement field under the simplifying transverse wave assumptions.

\subsubsection{Phase Unwrapping}
Oscillations with large amplitudes can cause $2\pi$ phase jumps that can hinder the estimation of the SWS. As in the original work (see \cite{hirsch2017magnetic, tzschatzsch2016tomoelastography}), we tackle these phase wraps by Laplacian-based unwrapping \cite{schofield2003fast}.

\subsubsection{Temporal Fast Fourier Transform (FFT)}
\label{subsubsec:FFT}
To isolate the fundamental harmonics of the wave corresponding to the excitation frequency used during MRE acquisition, we compute the FFT of the image phase $\boldsymbol{\hat{\varphi}} := \mathbf{F}_t(\boldsymbol{\varphi})$ and consider only the harmonic frequency, which we denote by $\boldsymbol{\bar\varphi} \in \mathbb{C}^{N}$.

\subsubsection*{Remark}
In general, in a continuous setting, a harmonic shear wave can be modeled as a complex-valued plane wave $\mathbf{u}(\mathbf{r}, t) = \mathbf{u}_0 \mathrm{e}^{\mathrm{i}(\mathbf{k} \cdot \mathbf{r} - \omega t)}$, where the complex wave vector $\mathbf{k}$ includes both propagation and attenuation, and $\omega$ is the angular frequency of the vibration. Here, we focus on estimating $\operatorname{Re}(\mathbf{k})$, which determines the wave speed. The real part of the phase of the planar shear wave (as encoded by the MEGs and captured in the MR image phase) is given by $\varphi(\mathbf{r}, t) = \operatorname{Re}(\mathbf{k}) \cdot \mathbf{r} - \omega t.$
Applying the temporal FFT and taking the phase gradient at the fundamental harmonic provides an estimate of the real part of the wave vector $\nabla_{\mathbf{r}} \text{arg}(\hat{\varphi}(\mathbf{r})) = \operatorname{Re}(\mathbf{k}(\mathbf{r}))$.

\subsubsection{Directional Filtering and Wave Vector Estimation}
For each of the $N_{\mathrm{MEG}}=3$ MEG directions $m\in \{m_x,m_y,m_z\}$ and $N_{\mathrm{freq}}$ frequencies $f\in \{f_1, \dots, f_{N_{\mathrm{freq}}}\}$, the wavefield $\boldsymbol{\bar{\varphi}}^{m,f}$ consists of superimposed waves. To isolate wavefronts propagating in several directions, the wavefield is filtered by $\bar{\boldsymbol{\varphi}}_h^{m,f} := \mathbf{F}^\herm \mathbf{H}_h \mathbf{F} \bar{\boldsymbol{\varphi}}^{m,f}$, with $\mathbf{H}_h$ being the linear operators associated with the directional filters $h$ and $\mathbf{F}$ a 2D FFT. Then a full wave vector $\mathbf{k}^{m,f,h}$ can be calculated for all $m$, $f$, and $h$ of the discretized filtered wavefields $\bar{\boldsymbol{\varphi}}_h^{m,f}$. To account for geometric attenuation, each filtered wave component is normalized, and the magnitude of the real wave vector in each direction is componentwise given by the phase-gradient:
\begin{equation}
    \|\operatorname{Re}(\mathbf{k}^{m,f,h}[l])\|_2 = \left\| \nabla_{\mathbf{r}}\,\text{arg} \left( \frac{\bar{\boldsymbol{\varphi}}_h^{m,f}[l]}{|\bar{\boldsymbol{\varphi}}_h^{m,f}[l]|} \right) \right\|_2.
\end{equation}

\subsubsection{Weighted Inversion}
Finally, the SWS $\mathbf{c}$ is computed from the real part of the estimated wave vectors. For voxel $l\in \{1,\ldots,N \}$, a weighted averaging is applied to combine multiple MEG directions and frequency components. Each measurement is weighted by 
$\mathbf{w}_{m,f,h}[l] = |\boldsymbol{\bar{\varphi}}_h^{m,f}[l]|^4$ \cite{hirsch2017magnetic}, which favors higher signal amplitudes. The SWS reads
\begin{align}\label{eq:kmdev_inversion}
    \mathbf{c}[l] = \Bigg(\frac{\sum_{m,f,h} \frac{ \|\operatorname{Re}(\mathbf{k}^{m,f,h}[l])\|_2}{2\pi f} \mathbf{w}_{m,f,h}[l]}{\sum_{m,f,h} \mathbf{w}_{m,f,h}[l]}\Bigg)^{-1} =: \\ \bigg(k\text{-MDEV}\Big( \big\{\mathbf{x}_{\theta}^{m,f} \big\}_{m,f}\Big)\bigg)[l],
\end{align}
where each image $\mathbf{x}_{\theta}^{m,f}$ is given by the CNN-based reconstruction $\mathrm{Rec}_{\theta}(\mathbf{y}^{m,f})$ with $\mathbf{y}^{m,f}$ denoting the $k$-space data for the MEG $m$ and frequency $f$. This yields spatially resolved SWS maps, which are used as the primary tissue parameter in this work. Additionally, it should be noted that due to GPU memory limitations in our end-to-end pipeline as described in \ref{sec:experiments}, we retain frequency-resolved SWS maps using only one image per MEG rather than using $N_{\mathrm{freq}}$. The entire $k$-MDEV pipeline (shown in Fig. $\ref{fig:Reco}$) is based on \cite{tzschatzsch2016tomoelastography, hirsch2017magnetic} and was implemented in \texttt{PyTorch} to allow for gradient backpropagation during joint image reconstruction and SWS estimation.

%% file: sections/4_experiments.tex
\section{Experiments} \label{sec:experiments}
We evaluated our proposed method on a brain MRE imaging problem. The experimental setup and the dataset are described in the following. The code for the image reconstruction pipeline is available at \url{https://github.com/Stef-Martin/E2E-MRE}. 
All operators were implemented using our in-house developed \texttt{PyTorch}-compatible Python package \texttt{MRpro} \cite{mrpro}.

\subsection{Data}
In total, in vivo MRE was performed in 13 healthy volunteers without a history of neurological disease (10M, 3F; mean age ± SD: 28.6 ± 8.6 years). Data from 12 of 13 subjects were used in the retrospective study, as detailed in Subsection \ref{retr_us_data}. In addition, another subject was scanned for the prospective study; see Subsection \ref{sec:meth_real_data} for acquisition details. The study was approved by the ethics committee of Charité – Universitätsmedizin Berlin in accordance with the Declaration of Helsinki. Every participant gave oral and written informed consent. The data were acquired using a clinical 3T MRI scanner (Magnetom Lumina, Siemens Healthineers, Erlangen, Germany) equipped with a 32-channel excite/receive head coil. 
Shear waves were induced using a cradle actuator system placed beneath the head and driven by compressed air at $N_{\mathrm{freq}}=4$ consecutive frequencies ($20$, $25$, $30$, and $35\,\text{Hz}$), with vibration initiated $2\,\text{s}$ before acquisition to ensure a steady state of harmonic motion. Motion was encoded into the phase of the complex MRE signal using trapezoidal, flow-compensated MEGs \cite{dittmann2016vivo} with an amplitude of $34\,\text{mT/m}$, a slew rate of $125\,\text{mT/m/ms}$, and a duration of $26.48\,\text{ms}$, resulting in motion-encoding efficiencies of $69.48$ (20 Hz), $98.61$ (25 Hz), $125.10$ (30 Hz), and $145.80\,\text{rad/mm}$ (35 Hz). The MEGs were consecutively applied along $N_{\mathrm{MEG}}=3$ directions \cite{herthum2022cerebral} (read, phase, and slice) sequentially encoding three-dimensional displacement fields. Data were acquired at $N_{\mathrm{freq}}=4$ frequencies with $N_{\mathrm{time}}=8$ vibration phase offsets equally spaced over a vibration period. 

In the retrospective study, Cartesian spin-echo MR images served as ground-truth for simulating spiral acquisitions according to model \eqref{eq:forward_model}, with trajectories incrementally rotated by the golden angle (GA, $137.5^\circ$). In the prospective study, spirals (also rotated by the GA) were acquired directly. To optimize $k$-space coverage, all spirals acquired for the different $N_{\mathrm{time}}$, $N_{\mathrm{freq}}$, and $N_{\mathrm{MEG}}$ were further rotated by the GA, ensuring complementary information collection in both spatial and temporal dimensions.

\subsubsection{Retrospectively Undersampled Data} 
\label{retr_us_data}
Brain data from 12 volunteers were acquired with a Cartesian spin-echo MRE sequence\cite{herthum2022cerebral} with a field of view (FOV) of $202\,\text{mm}\,\times\, 202\,\text{mm}\,\times\,80\,\text{mm}$ and a resolution of $1.6\,\text{mm}\,\times\,1.6\,\text{mm}\,\times\,2\,\text{mm}$. The scan-time per subject was $451\,\text{s}$ with an echo time (TE) of $70\,\text{ms}$ and a repetition time (TR) of $4700\,\text{ms}$. These Cartesian reference data were acquired with an image size of $N_x \times N_y \times N_{\mathrm{slices}} = 126\times126\times40$.
Based on this, we retrospectively simulated accelerated MRE data acquisition according to \eqref{eq:forward_model} using $N_{\mathrm{spirals}}\in\{1,2,3\}$ spiral trajectories with 863 $k$-space points each (identical to those employed for in vivo scans in \ref{sec:meth_real_data}), yielding undersampling factors of 19, 9.5, and 6.3, respectively. Note that the forward operator $\mathbf{A}$ in \eqref{eq:forward_model} inherently depends on the CSMs, i.e., $\mathbf{A} = \mathbf{A}(\mathbf{C})$, which must be estimated before approaching the reconstruction. To mimic a realistic situation found in clinical practice, $k$-space data were retrospectively simulated according to \eqref{eq:forward_model} with $\mathbf{A}(\mathbf{C}_{\mathrm{true}})$, where $\mathbf{C}_{\mathrm{true}}$ were computed from spiral MRE data acquired with 25 spirals and 863 $k$-space points per spiral using the Walsh algorithm. An estimate $\tilde{\mathbf{C}}$ of these CSMs was obtained from retrospectively simulated Cartesian calibration $k$-space data, which were generated using the same CSMs, but only a small calibration set of $24 \times 24$ $k$-space points, thereby introducing realistic calibration uncertainties in the form of a model-mismatch, i.e.\,  $\mathbf{A}(\mathbf{C}_{\mathrm{true}}) \neq \mathbf{A}(\tilde{\mathbf{C}})$. The operator $\mathbf{A}(\tilde{\mathbf{C}})$ was used in the image reconstruction, see \ref{subsec:image_reconstruction}. The DCF $\mathbf{W}$ was derived using the Voronoi method. The dataset comprised $5760$ samples -- 12 subjects $\times$ 40 slices $\times$ 4 frequencies $\times$ 3 MEGs -- each associated with 8 distinct vibration phase offsets. The dataset was partitioned into training, validation, and test subsets using 8, 1, and 3 subjects, respectively, which correspond to $3840$, $480$, and $1440$ samples.

\subsubsection{Prospective In Vivo Study}
\label{sec:meth_real_data}
To evaluate our pipeline on in vivo data, we designed an MRE sequence using PyPulseq \cite{ravi2019pypulseq} and acquired data from a single volunteer. The acquisition parameters were set to an FOV of $220\,\text{mm}\,\times\,220\,\text{mm}\,\times\,5\,\text{mm}$, with an in-plane resolution of $1.7\,\text{mm}$, a flip angle of $10^\circ$, a TE of $21\,\text{ms}$, and a TR of $56\,\text{ms}$. Data were acquired using 25 spirals, each consisting of 863 $k$-space points, of which only a subset of $N_{\mathrm{spirals}}\in\{1,2,3\}$ was used in the model. CSMs were estimated from a reference dataset acquired using 10 spiral interleaves with a total acquisition time of 2.5 s. Prior to reconstruction with any method, the considered $k$-space data were normalized by their standard deviation.

\subsection{Network Architectures} \label{subsec:NN_arch}
As the CNN $u_{\theta}$ (see \eqref{eq:net_application}) employed in the $\mathrm{Rec}_{\theta}$ block, we use the XTYT network \cite{kofler2019spatio} -- specifically designed to reconstruct spatiotemporal slices. For the ablation study, we compared models with different numbers of unrolled iterations $T$ using XTYT blocks of different sizes. Both configurations employed three encoding and decoding stages with two convolutions per stage. For our proposed method with $T=6$, an XTYT-block with about 2.1 million parameters was used, whereas for the ablation study and $T=1$, the network used around 8.3 million trainable parameters, allowing for a more meaningful comparison of the approaches, see \cite{kofler2022more}.

\subsection{Loss Functions and Training} \label{subsec:loss_train}
Here, we consider two primary training strategies. 
Let $\mathbf{x}_{\theta}:=\mathrm{Rec}_{\theta}(\mathbf{y})$ and $\mathbf{c}_{\theta}:=k$-$\mathrm{MDEV}(\mathbf{x}_{\theta})$ denote the image reconstructed with the network described in Subsection \ref{subsec:image_reconstruction} and the SWS maps estimated by $k$-MDEV described in Subsection \ref{subsec:kmdev}, respectively, with $\mathbf{x}_{\text{target}}$ and $\mathbf{c}_{\text{target}}$ representing their corresponding targets. Both depend on the trainable parameters $\theta$, which, in turn, are determined by the network architecture as described in \ref{subsec:NN_arch}. Importantly, although only the image reconstruction block contains learnable parameters, we have the flexibility to choose the target in our network optimization to be either the complex-valued MR image or the SWS maps (or a combination thereof). By doing so, we can tailor the reconstruction network to the ultimate specific task, i.e.\ the estimation of the SWS maps. Here, we consider a loss function that combines contributions from qualitative MR images and SWS maps, given by
\begin{equation}\mathcal{L}_{\theta}^\text{Total}=\alpha\mathcal{L}_{\theta}^\text{Images}+(1-\alpha)\mathcal{L}_{\theta}^{\text{SWS}},
    \label{eq:loss_tot}
\end{equation}
where $\mathcal{L}_{\theta}^\text{Images}$ and $\mathcal{L}_{\theta}^{\text{SWS}}$ are defined as
\begin{equation}
    \begin{aligned}
        \mathcal{L}_{\theta}^\text{Images}(\mathbf{x}_{\theta},\mathbf{x}_{\text{target}})&= \mathcal{L}^{\perp}(\mathbf{x}_{\theta},\mathbf{x}_{\text{target}})+\mathrm{MAE}(|\mathbf{x}_{\theta}|,|\mathbf{x}_{\text{target}}|),\\
    \end{aligned}
    \label{eq:loss_image}
\end{equation}
and
\begin{equation}
    \begin{aligned}
        \mathcal{L}_{\theta}^{\text{SWS}}(\mathbf{c}_{\theta},&\mathbf{c}_{\text{target}})
        = \frac{1}{2} \big(\gamma \cdot \mathrm{RMSE}(\mathbf{c}_{\theta},\mathbf{c}_{\text{target}}) \\
        & + (1 - \mathrm{SSIM}(\mathbf{c}_{\theta},\mathbf{c}_{\text{target}})) \big),
    \end{aligned}
    \label{eq:loss_SWS}
\end{equation}
with the deviation between image estimate and target being the perpendicular loss $\mathcal{L}^{\perp}$ %
as given by \cite{terpstra2022loss} and with $\mathrm{MAE}$ denoting the mean absolute error. In \eqref{eq:loss_SWS}, the root mean squared error ($\mathrm{RMSE}$ in m/s) quantifies the average magnitude of errors between SWS maps, while the structural similarity index ($\mathrm{SSIM}$, unitless) measures their perceptual similarity --  both metrics are on a comparable scale. A weight of $\gamma=2$ has been assigned to the $\mathrm{RMSE}$ term in \eqref{eq:loss_SWS}. Note that while we use RMSE for training, we additionally report the normalized RMSE (NRMSE) for evaluation; that is, the RMSE normalized by the reference image/SWS map energy, which provides a dimensionless and scale-invariant error measure.

\subsubsection{Pre-Training}\label{sec:pretrain}
The most straightforward approach is to train the network directly with fully sampled complex MRI images serving as the target\cite{martinmr}. In this method, $k$-MDEV is treated as an independent module and does not influence the training process, as visualized in Fig. \ref{fig:Reco}, yellow box. This corresponds to setting $\alpha=1$ in \eqref{eq:loss_tot}. AdamW \cite{loshchilov2017fixing} was chosen as the optimizer with a learning rate of $10^{-4}$ and a weight decay of $10^{-2}$. After each epoch, the SWS maps were calculated from the image estimate and compared to the corresponding reference SWS maps using \eqref{eq:loss_SWS}. This loss was tracked over the whole pre-training process. The initialization block (see purple block in Fig. \ref{fig:Reco}) uses 24 iterations to approximate the Moore-Penrose pseudo-inverse solution.
Pre-training required approximately 1.5 hours per epoch, totaling 3 days for 50 epochs of training (on a single Nvidia A6000) with GPU requirements of 13.2 GB.

\subsubsection{End-to-End Fine-Tuning}\label{sec:end2end-training}
Our final proposed end-to-end approach, displayed in Fig. \ref{fig:Reco}, green box, integrates the entire $k$-MDEV module into the model as a post-processing step applied to the network output (the estimated MR images) and used in the training process. The SWS loss \eqref{eq:loss_SWS} can then be computed by comparing the resulting SWS maps to the target SWS maps, which were obtained by applying $k$-MDEV to the reference MR images.
End-to-end training corresponds to setting $0<\alpha<1$ in \eqref{eq:loss_tot}. In this subsequent task-specific fine-tuning stage, the pre-trained weights from \ref{sec:pretrain} are further adapted, shifting the optimization towards the final quantitative objective of accurate SWS estimation, while maintaining clinical interpretability of the qualitative images due to the data-consistent reconstruction framework and the choice of $\alpha>0$, in our case $\alpha=0.1$. Since the network is already pre-trained according to \ref{sec:pretrain}, a very small initial learning rate of $5\times10^{-6}$ was chosen (decreasing linearly to $10^{-7}$ over the 50 epochs) in the AdamW optimizer with a weight decay of $10^{-2}$. The learning rate of the regularization parameter $\lambda$ was chosen to be $3\times10^{-2}$. We included MR images from multiple MEG directions for SWS calculation, as this yields more reliable estimates. Frequency-resolved maps, however, provide a sufficiently accurate basis for training. Their discrepancy from weighted multi-frequency maps according to \eqref{eq:kmdev_inversion} is 0.19 in terms of \eqref{eq:loss_SWS}. While this difference is not negligible, weighted maps can be reconstructed retrospectively from frequency-resolved ones using \eqref{eq:kmdev_inversion}. Furthermore, since the weighting depends on the magnitude of the SWS while our loss \eqref{eq:loss_SWS} measures RMSE, the improvements achieved on the frequency-resolved maps extend directly to the weighted maps.
To minimize the influence of artifacts in the SWS maps outside the tissue, a binary brain mask was applied during loss calculation. The application of this mask is crucial, as it ensures that the network focuses primarily on the brain region, preventing it from merely removing artifacts in the non-brain regions. Fine-tuning required approximately 1.5–2 hours per epoch, corresponding to about 4 days for 50 epochs on a single Nvidia A6000 GPU, with 35.4 GB memory usage. Overall, pre-training and fine-tuning together amounted to roughly one week of computation time. After training, the feed‑forward time during inference to compute an SWS map from undersampled $k$-space data with 8 vibration phase offsets is approximately 0.75 s.

\subsection{Methods of Comparison} \label{subsec:comparison}
For comparison, we included two additional reconstruction methods. First, an iterative SENSE \cite{pruessmann2001advances} reconstruction was employed as a conventional parallel imaging approach. Second, we implemented a TV-regularized reconstruction promoting temporal gradient sparsity of the images, following \cite{caballero2014dictionary}. The problem was solved using a variable splitting scheme that alternates between CG and PDHG \cite{chambolle2011first}, which was implemented with \texttt{MRpro}\cite{mrpro}.

\subsection{Different Acceleration Factors}\label{subsec:experiments}
To address varying trajectories and the resulting undersampling artifacts, the model was trained using trajectories containing one, two, or three spiral arms with 863 $k$-space points each. These correspond to undersampling factors of approximately 19, 9.5, and 6.3, respectively. The trajectories were randomly selected for retrospective undersampling during training, enabling the network to handle different acceleration factors within a unified framework. The trained model was evaluated on simulated data and applied to real acquisitions to demonstrate its robustness and in vivo applicability across varying sampling conditions. Additionally, its performance was evaluated against our comparison methods as introduced in \ref{subsec:comparison}.

%% file: sections/5_results.tex
\section{Results} \label{sec:results}

\subsection{Pre-Training vs. End-to-End Fine-Tuning}\label{subsec:results_pretr_e2e}
As outlined in Section \ref{subsec:loss_train}, the training procedure comprises two stages: an initial pre-training phase, where the model is trained using qualitative MR images as targets, followed by a fine-tuning phase that leverages the pre-trained weights and uses the SWS maps as targets. Here, we illustrate the impact of each stage by comparing the model performance before and after fine-tuning.

\subsubsection{Pre-Training}
\begin{figure}
    \centering
    \includegraphics[width=\linewidth]{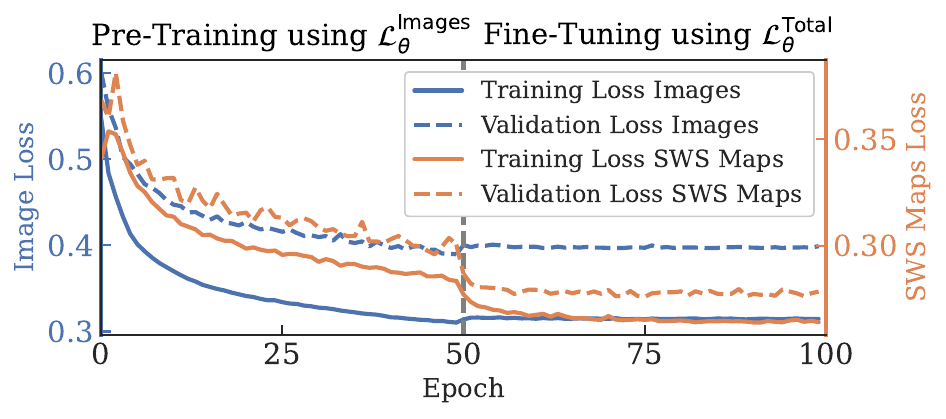}
    \caption{Training and validation loss of complex MR images and SWS maps over 100 epochs with $T=6$. Perpendicular loss \eqref{eq:loss_image} was used for images, RMSE combined with SSIM for SWS maps \eqref{eq:loss_SWS}. The first 50 epochs correspond to pre-training with the qualitative images as reference, followed by 50 epochs of fine-tuning with $\alpha=0.1$ according to \eqref{eq:loss_tot}.}
    \label{fig:E2E_loss}
\end{figure}

The left section of Fig. \ref{fig:E2E_loss} (up to the vertical dashed line) shows the training and validation loss curves for both reconstructed images and SWS maps throughout the 50-epoch pre-training phase. Blue lines indicate losses for estimated images relative to reference images, while orange lines represent SWS map losses, which are not used in training but monitored for comparison. 
Both losses decrease steadily, indicating improved model performance for both tasks, with convergence observed near 50 epochs. The image reconstruction loss drops significantly to ~56\% of its initial value, whereas the SWS map loss decreases more modestly to 83\%. While these losses are not directly comparable due to differing loss functions, the observed discrepancy remains noteworthy.

\subsubsection{End-to-End Fine-Tuning}
Building upon the converged pre-trained network, the deviation from the SWS map targets can be further reduced through additional fine-tuning within an end-to-end optimization framework. The right segment of Fig. \ref{fig:E2E_loss}, beyond the vertical dashed line, shows the training and validation loss curves over an additional 50 epochs, with optimization focused on minimizing SWS loss. This fine-tuning stage yields a further reduction of approximately 5\% in the validation loss of the SWS maps. Concurrently, the image reconstruction loss is monitored, revealing a slight increase as the model prioritizes minimizing the SWS error. Given that SWS maps derived from the combination of multiple MEGs yield more accurate representations compared to those obtained from a single MEG, the network was trained using data from all acquired MEGs. However, to maintain consistency and comparability with the image pre-training phase, the loss computed for each individual MEG direction is presented.

\begin{figure}
    \centering
    \includegraphics[width=0.95\linewidth]{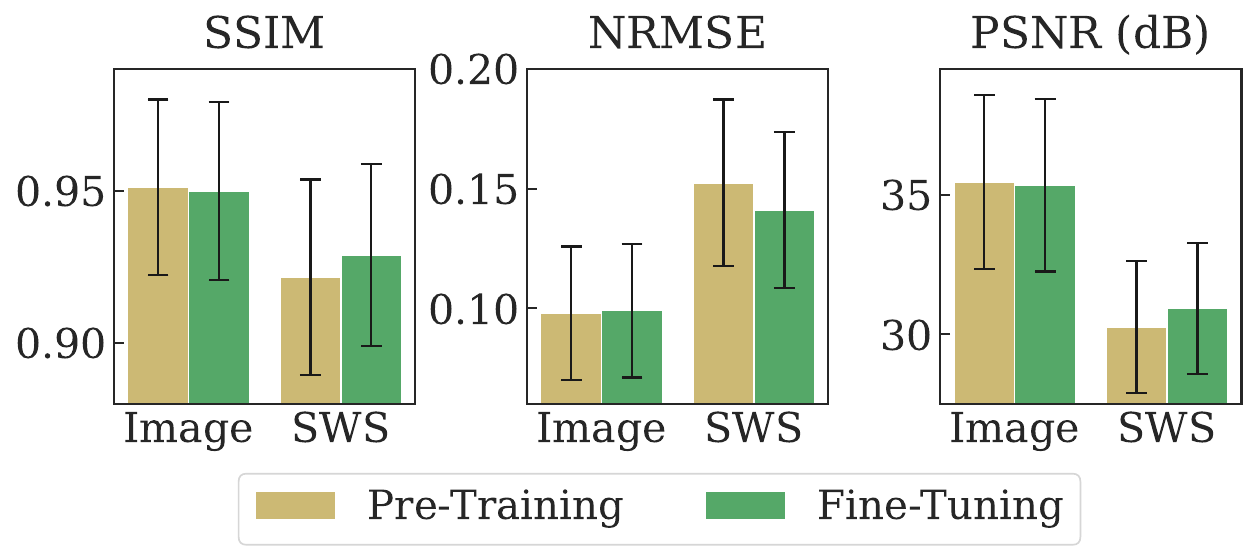}
    \caption{SSIM, NRMSE, and PSNR computed on the test set with pre-trained and fine-tuned model ($T=6$, $\alpha=0.1$) applied. The metrics were calculated on qualitative images and SWS with $N_\mathrm{meg}=3$. Except for the qualitative images, the fine-tuned model outperformed the pre-trained one.}
    \label{fig:Pretr_vs_E2E_metrics}
\end{figure}

Fig. \ref{fig:Pretr_vs_E2E_metrics} shows the improvements in SWS maps, illustrating SSIM, NRMSE, and PSNR for qualitative images and SWS maps with $N_\mathrm{meg}=3$ using pre-trained and fine-tuned models applied to the test set undersampled with 1 spiral arm. A negligible model performance decline is observed exclusively for the qualitative images. In contrast, the metrics calculated for the SWS maps reflect improved model performance following fine-tuning.

\subsection{Comparison to Standard Methods} \label{subsubsec:results_comparison}
\begin{figure}
    \centering
    \includegraphics[width=0.95\linewidth]{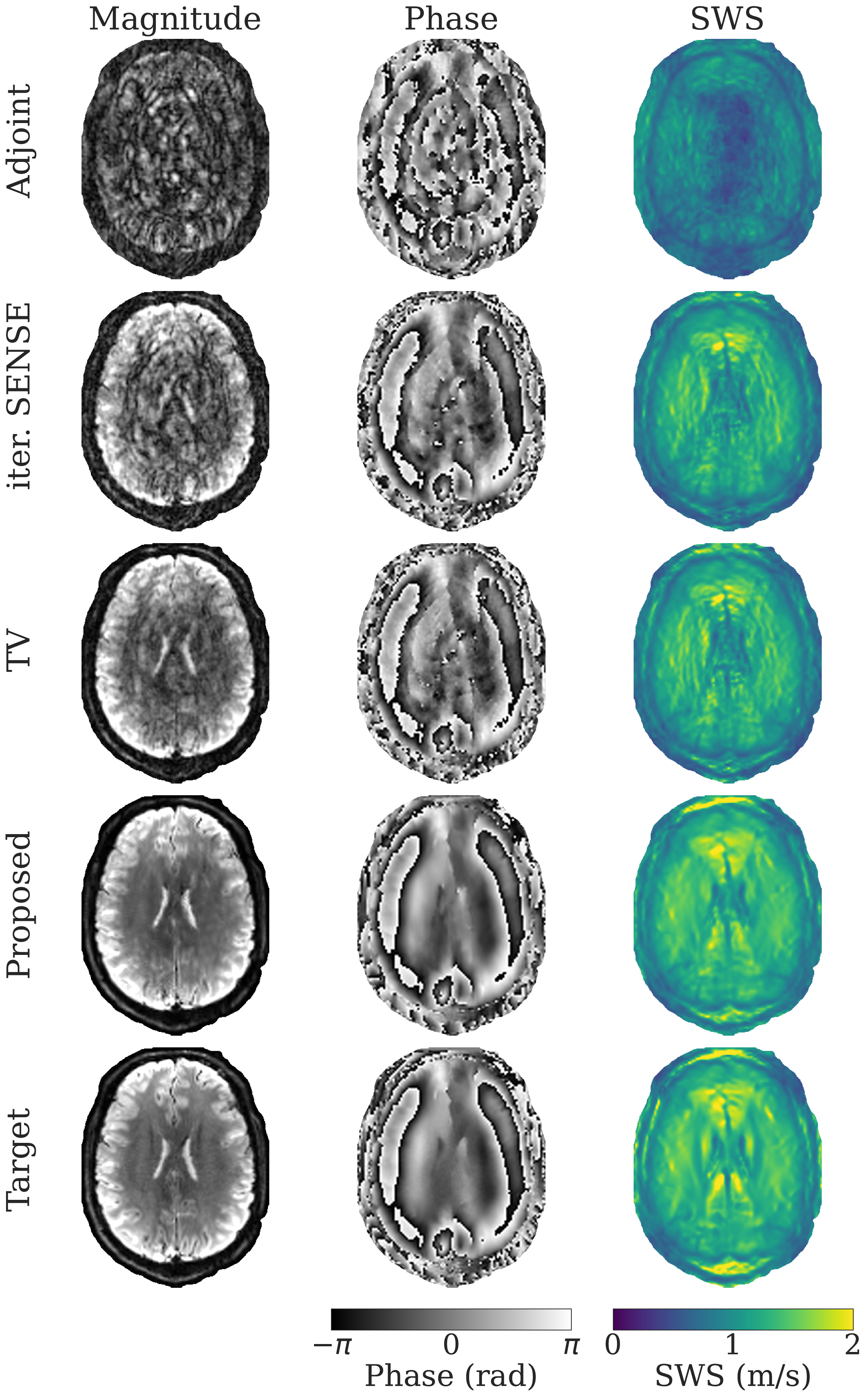}
    \caption{Magnitude images (left), phase images (middle), and resulting SWS maps (right) from reconstructions using various methods on retrospectively undersampled single-spiral data. From top to bottom: Adjoint, iterative SENSE \cite{pruessmann2001advances}, the TV minimization-based comparison method, and the proposed method. Fully sampled data (bottom row) serves as reference.}
    \label{fig:comp_methods}
\end{figure}

We assess the performance of our proposed end-to-end approach in comparison to conventional reconstruction techniques, including iterative SENSE and our TV-regularized comparison method. Fig. \ref{fig:comp_methods} presents magnitude and phase images of a representative slice, along with the SWS map derived from three MEGs, illustrating the outcomes of different reconstruction methods. Specifically, we compare a straightforward adjoint reconstruction, the pseudo-inverse, the comparison method, and our proposed method against the reference target images. Our approach achieves superior reconstruction quality, delivering clearer and more accurate results than conventional techniques for both qualitative and quantitative images. Table \ref{tab:real_data_metrics} summarizes the quantitative results for SSIM, NRMSE, and PSNR computed on the masked SWS maps, showing that our approach substantially outperforms all compared methods. While the TV-based comparison method produces clear improvements in magnitude images as seen in Fig. \ref{fig:comp_methods}, the masked SWS maps exhibit only minor enhancements relative to iterative SENSE.

\begin{table}[ht]
\renewcommand{\arraystretch}{1.3}
\centering
\begin{tabular}{|l|c|ccc|}
\hline
\textbf{Spirals} & \textbf{Method} & \textbf{SSIM} & \textbf{NRMSE} & \textbf{PSNR (dB)} \\
\hline
\multirow{4}{*}{1}
    & Adjoint     & 0.75 $\pm$ 0.072 & 0.43 $\pm$ 0.054 & 21.10 $\pm$ 2.96 \\
    & it. SENSE & 0.83 $\pm$ 0.051 & 0.23 $\pm$ 0.040 & 26.79 $\pm$ 2.69 \\
    & TV  & 0.87 $\pm$ 0.049 & 0.20  $\pm$ 0.041 & 27.81 $\pm$ 2.58 \\
    & Proposed    & \bf{0.93 $\pm$ 0.030} & \bf{0.14 $\pm$} 0.033 & \bf{30.92 $\pm$ 2.34} \\
\hline
\multirow{4}{*}{2}
    & Adjoint     & 0.83 $\pm$ 0.055 & 0.26 $\pm$ 0.030 & 25.37 $\pm$ 2.88 \\
    & it. SENSE & 0.92 $\pm$ 0.035 & 0.16 $\pm$ 0.036 & 30.10 $\pm$ 2.65 \\
    & TV  & 0.93 $\pm$ 0.035 & 0.15 $\pm$ 0.036 & 30.61 $\pm$ 2.64 \\
    & Proposed    & \bf{0.96 $\pm$ 0.019}  & \bf{0.11 $\pm$ 0.031}  & \bf{33.03 $\pm$ 2.39}  \\
\hline
\multirow{4}{*}{3}
    & Adjoint     & 0.88 $\pm$ 0.041 & 0.20 $\pm$ 0.027 & 27.81 $\pm$ 2.61 \\
    & it. SENSE & 0.94 $\pm$ 0.027 & 0.14 $\pm$ 0.036 & 31.30 $\pm$ 2.69 \\
    & TV  & 0.95 $\pm$ 0.026 & 0.13 $\pm$ 0.037 & 31.83 $\pm$ 2.67 \\
    & Proposed    & \bf{0.97 $\pm$ 0.015}  & \bf{0.10 $\pm$ 0.031}  & \bf{34.02 $\pm$ 2.44}  \\
\hline
\end{tabular}
\caption{Comparison of quantitative metrics (SSIM, NRMSE, PSNR) computed from SWS maps reconstructed using different methods, applied to simulated undersampled data with 1, 2, and 3 spirals. Masked metrics are shown, using the brain mask as explained in \ref{sec:end2end-training}. Best values shown in bold.}
\label{tab:real_data_metrics}
\end{table}

\subsection{Different Undersampling Factors}\label{subsubsec:results_123}
Table \ref{tab:real_data_metrics} presents SWS map evaluation results for $N_\mathrm{meg}=3$ using SSIM, NRMSE, and PSNR metrics on the test set. As expected, performance improves across all metrics with an increasing number of spirals.  A substantial gain is observed when moving from one to two spirals. These trends are consistent with the results of the reconstructed qualitative images.

\subsection{Ablation Study}\label{subsec:ablation}
\begin{figure}[h]
    \centering
    \includegraphics[width=\linewidth]{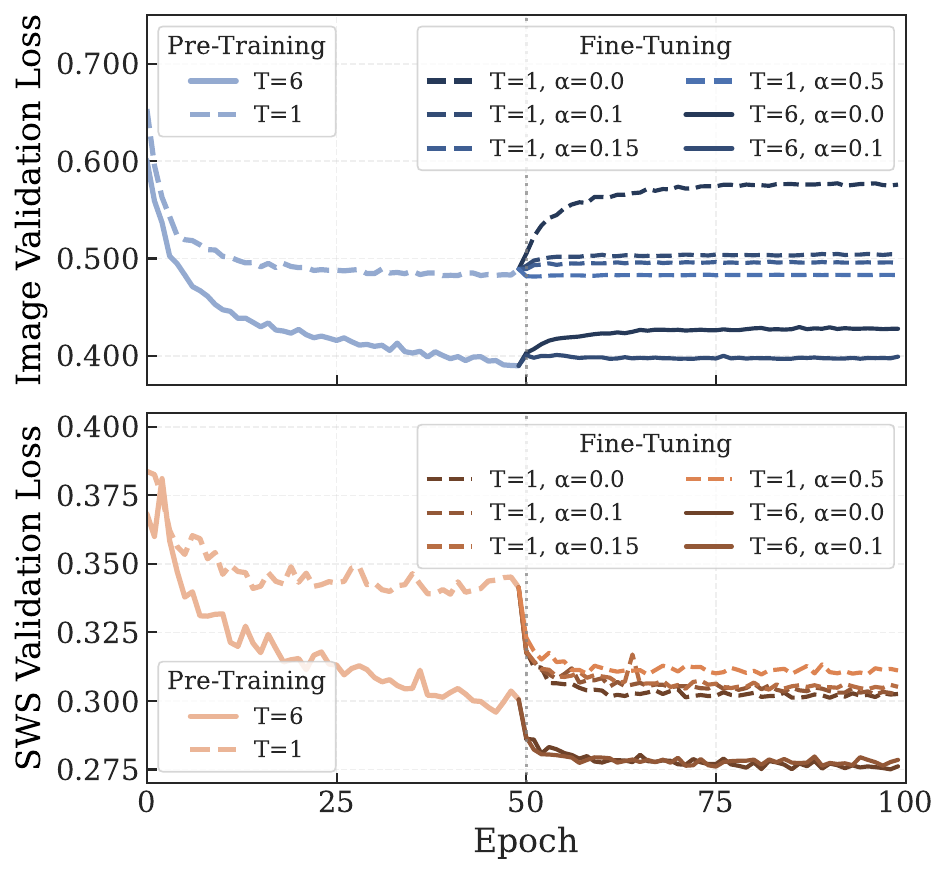}
    \caption{Comparison between different numbers of unrolled iterations ($T=1$ vs $T=6$) as well as different hyper-parameters $\alpha \geq 0$ for the loss function \eqref{eq:loss_tot} used for fine-tuning the entire end-to-end trainable network. Employing a larger $T$ yields a lower validation error. Additionally, reducing the effect of catastrophic forgetting with respect to the task of image reconstruction is possible by setting $\alpha>0$. The proposed parameter combination $T=6$ and $\alpha=0.1$ yields a good tradeoff between image- and SWS-accuracy.
    }
    \label{fig:comparison_T_alpha}
\end{figure}

To evaluate the impact of key design choices of our model, we conducted an ablation study examining two critical parameters: the number of unrolled iterations $T$ and the loss weighting coefficient $\alpha$, which balances the combined loss \eqref{eq:loss_tot}. Our analysis reveals that increasing $T=1$ to $T=6$ improves SWS estimation quality by 18\% when keeping $\alpha$ constant. Regarding the loss weighting, we observe a trade-off between SWS accuracy and image quality: setting $\alpha$ close to zero yields improved SWS metrics but results in substantial degradation of qualitative image reconstruction (12\% when moving from $\alpha=0.15$ to $\alpha=0$), whereas higher $\alpha$ values preserve image fidelity at the expense of SWS accuracy. Fig. \ref{fig:comparison_T_alpha} and Table \ref{tab:comparison_T_alpha} summarize the described results.

\begin{table}[ht]
\renewcommand{\arraystretch}{1.3}
\centering
\begin{tabular}{|c|c|c|ccc|}
\hline
 & T & $\alpha$ & \textbf{SSIM} & \textbf{NRMSE} & \textbf{PSNR (dB)} \\
\hline
\multirow{6}{*}{\rotatebox{90}{Image}}
& 1 & 0.00 & 0.90 $\pm$ 0.041 & 0.16 $\pm$ 0.031 & 31.45 $\pm$ 2.64 \\
& 1 & 0.10 & 0.91 $\pm$ 0.042 & 0.14 $\pm$ 0.033 & 31.69 $\pm$ 2.74 \\
& 1 & 0.15 & 0.91 $\pm$ 0.041 & 0.14 $\pm$ 0.032 & 31.75 $\pm$ 2.75 \\
& 1 & 0.50 & 0.91 $\pm$ 0.041 & 0.14 $\pm$ 0.033 & 31.90 $\pm$ 2.75 \\
& 6 & 0.00 & \bf{0.95 $\pm$ 0.030} & 0.10 $\pm$ 0.029 & 34.97 $\pm$ 3.07 \\
& 6 & 0.10 & \bf{0.95 $\pm$ 0.029} & \bf{0.09 $\pm$ 0.028} & \bf{35.34 $\pm$ 3.09} \\
\hline
\multirow{6}{*}{\rotatebox{90}{SWS}}
& 1 & 0.00 & 0.90 $\pm$ 0.039 & 0.17 $\pm$ 0.032 & 29.47 $\pm$ 2.38 \\
& 1 & 0.10 & 0.90 $\pm$ 0.040 & 0.17 $\pm$ 0.032 & 29.35 $\pm$ 2.40 \\
& 1 & 0.15 & 0.89 $\pm$ 0.040 & 0.17 $\pm$ 0.032 & 29.31 $\pm$ 2.41 \\
& 1 & 0.50 & 0.89 $\pm$ 0.041 & 0.18 $\pm$ 0.032 & 29.21 $\pm$ 2.42 \\
& 6 & 0.00 & \bf{0.93 $\pm$ 0.030} & \bf{0.14 $\pm$ 0.033} & \bf{30.92 $\pm$ 2.32} \\
& 6 & 0.10 & \bf{0.93 $\pm$ 0.030} & \bf{0.14 $\pm$ 0.033} & \bf{30.92 $\pm$ 2.34} \\
\hline
\end{tabular}
\caption{Reconstruction metrics for different $T$ and $\alpha$ configurations applied to the 19-fold undersampled test set. SWS maps are calculated for all MEGs and frequencies combined.}
\label{tab:comparison_T_alpha}
\end{table}

\subsection{In Vivo Data}\label{subsec:results_real_data}
\begin{figure}
    \centering
    \includegraphics[width=0.9\linewidth]{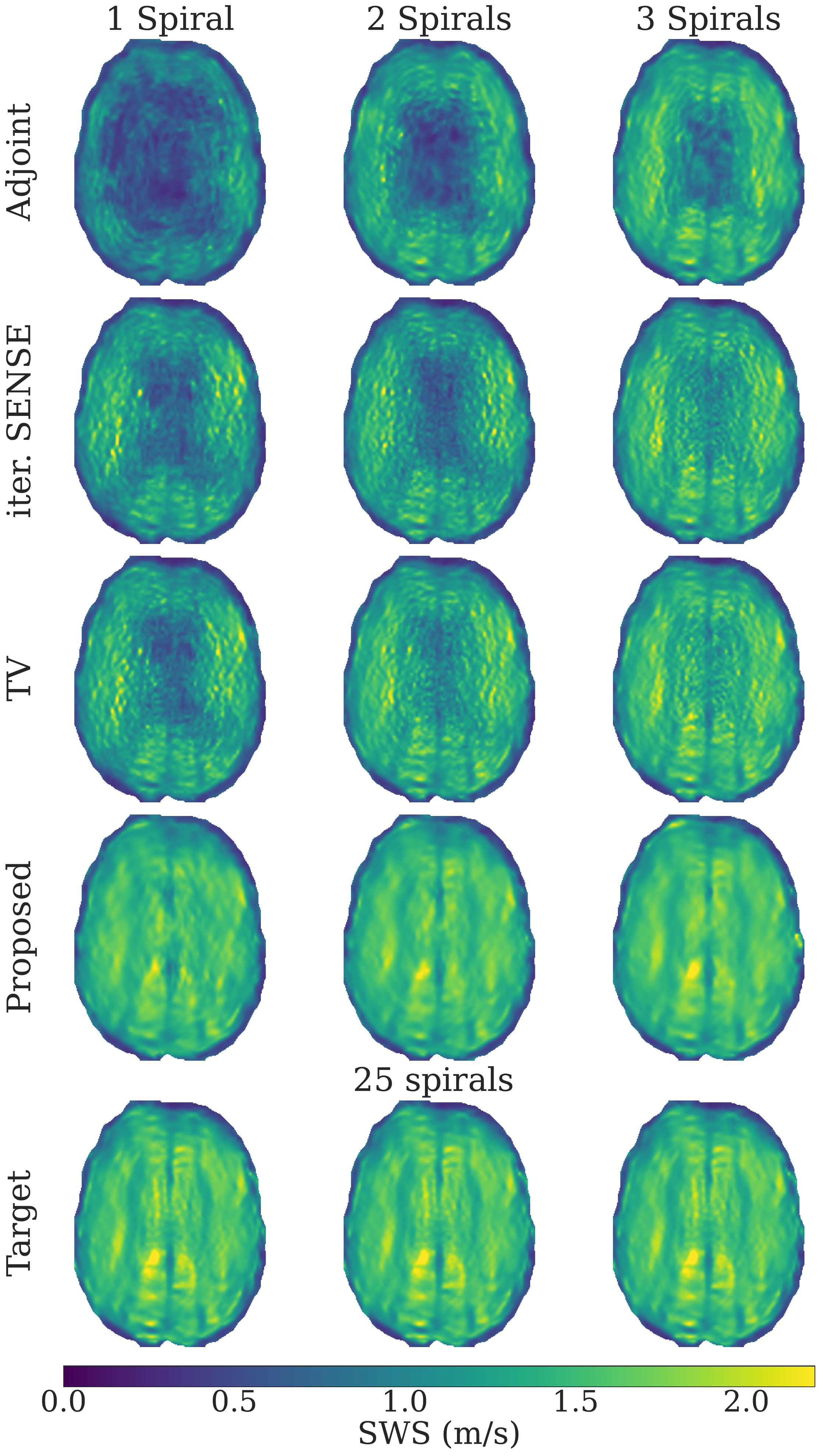}
    \caption{SWS maps resulting from the application of different methods to in vivo data acquired with one, two or three spirals compared to the target SWS maps of an acquisition with 25 spirals.}
    \label{fig:SWS_real}
\end{figure}

Fig. \ref{fig:SWS_real} displays the SWS maps obtained by applying the proposed method to the undersampled $k$-space data of the in vivo data for one to three spirals, compared to iterative SENSE, the TV-regularized comparison method, and the reference SWS maps obtained using 25 spirals. The SWS maps displayed were computed with all acquired $N_{\mathrm{meg}}=3$ MEGs, $N_{\mathrm{freq}}=4$ frequencies, and $N_{\mathrm{time}}=8$ time steps. The resulting SWS maps are sufficiently accurate even with a single spiral; however, they lack finer detail. For two and three spirals, they closely match the reference. 

%% file: sections/6_discussion.tex
\section{Discussion} \label{sec:discussion}
In this work, we presented a novel end-to-end approach for computing SWS maps directly from highly undersampled $k$-space data. In contrast to ultrasound elasticity imaging, where physics-informed neural networks estimate stiffness directly from measured wavefields\cite{yin2023swenet} -- which are subject to depth limitations and acoustic artifacts -- our MRE framework integrates physics-informed learning into the MR image reconstruction stage. Here, the SWS maps -- derived from an established MRE inversion -- guide the reconstruction toward high displacement fidelity under standardized slice-selective encoding and low SNR conditions. This approach naturally extends to 3D acquisitions beyond the 2D implementation presented in this work. This is, to the best of our knowledge, the first end-to-end trainable physics-informed reconstruction network in MRE that jointly performs image reconstruction and SWS estimation directly from $k$-space. We discuss key aspects of this framework in the following sections.

\subsection{Reconstruction Method Design}
Our proposed method integrates $k$-MDEV \cite{tzschatzsch2016tomoelastography} as a differentiable layer within a model-based deep learning approach for MR image reconstruction, following MoDL \cite{aggarwal2018modl}. We employ an XTYT architecture \cite{kofler2019spatio} to exploit temporal correlations in the phase (compared to magnitude in cardiac cine MRI) across vibration phase offsets, with loss functions\cite{terpstra2022loss} designed to preserve both magnitude \textit{and} phase fidelity critical for MRE. The sampling patterns vary temporally, enabling effective regularization by leveraging redundancies across both spatial and temporal dimensions. In doing so, the reconstruction problem remains relatively small, consisting of $N_{\mathrm{time}}$ 2D MR reconstruction problems that are solved simultaneously. This setup allows for the construction of unrolled methods such as the one considered here, where we used $T=6$ unrolling steps as defined by \eqref{eq:net_application} and \eqref{eq:dc_update}. Exploiting redundancies along different dimensions (e.g., the $z$-dimension across slices) could be an interesting alternative direction. Given that typically $N_{\mathrm{slices}} > N_{\mathrm{time}}$ ($N_{\mathrm{slices}}=40$ vs.\ $N_{\mathrm{time}}=8$ in our case), constructing unrolled networks with a larger number of outer iterations $T$ might be computationally prohibitive. Setting $T=1$ and opting for a much larger network $u_{\theta}$, as proposed in \cite{kofler2020neural} and \cite{kofler2021end}, and also studied here in the ablations study in \ref{subsec:ablation}, provides computational benefits and certain theoretical advantages (namely, convexity of the implicitly defined reconstruction problem). However, as demonstrated in our ablation study (see Fig. \ref{fig:comparison_T_alpha}), we find that choosing $T=6$ yields substantially improved reconstruction performance.
Regarding the choice of the loss function used for end-to-end fine-tuning, setting $\alpha=0$ leads to a very mild form of catastrophic forgetting in the network, as illustrated in Fig. \ref{fig:comparison_T_alpha}. In this case, the qualitative image produced by the model deteriorates noticeably -- compared to the pre-training results -- to enhance the quality of the estimated SWS maps. Although the data-consistent architecture of our model mostly mitigates this decrease in reconstruction quality, we observe that $\alpha=0.1$ represents an optimal trade-off: the image quality remains largely preserved, while the model continues to prioritize accurate SWS reconstruction without noticeable compromise (see also Table \ref{tab:comparison_T_alpha}, with $\alpha=0.15$ and $T=1$, SSIM for SWS maps decreases while image quality is mostly the same compared to $\alpha=0.1$ and $T=1$).

\subsection{Training Strategies}
For training, we explored two strategies visualized in Fig. \ref{fig:Reco}: using qualitative images alone versus using both SWS maps \textit{and} qualitative images as targets (see loss function in \eqref{eq:loss_tot}). The latter enables the reconstruction network $\mathrm{Rec}_{\theta}$ to learn representations that facilitate SWS estimation rather than solely optimizing image quality.
Our results indicate that, consistent with other studies, e.g.\ for brain T1-mapping in \cite{zimmermann2024pinqi}, the achievable performance of reconstruction networks for qMRI problems can be improved if they are additionally trained for the subsequent downstream task -- here, SWS quantification -- rather than separating image reconstruction problem and SWS estimation, as seen in Fig. \ref{fig:Pretr_vs_E2E_metrics}. The level of improvement is expected to vary between different qMRI problems and, although visible here, was less prominent than in \cite{zimmermann2024pinqi}. We suspect that this relates to the nature and level of ill-posedness of the considered problem. Future research could explore which problems benefit from this approach and to what extent.

\subsection{Reconstruction Performance}
Compared to conventional image reconstruction techniques such as iterative SENSE \cite{pruessmann2001advances} and TV minimization \cite{block2007undersampled}, our approach consistently produces more accurate qualitative images and SWS maps, as illustrated in Fig. \ref{fig:comp_methods}. 
Remarkably, all deep learning-based reconstruction approaches tested outperformed classical methods by a large margin, independent of network architecture and size or training strategy, highlighting the effectiveness of incorporating a CNN block within a regularized image reconstruction framework. 
Furthermore, conventional reconstruction techniques focus solely on producing qualitative images and do not consider the calculation of SWS maps. In contrast, our proposed end-to-end model integrates $k$-MDEV as a fixed, differentiable module, enabling direct estimation of SWS maps within the reconstruction framework. To assess the impact of this integration, we compared our pre-training approach -- formulated as a regularized image reconstruction model -- against the full end-to-end model. Our results demonstrate that the end-to-end approach outperforms the standalone reconstruction model, reducing SWS map deviation by 5\%. These findings suggest that jointly optimizing image reconstruction and SWS estimation within an end-to-end framework better preserves the quantitative information in the phase, which is essential for accurate SWS mapping. This indicates that by incorporating regularization that jointly considers both tasks, the model achieves more reliable and consistent results compared to methods that treat these tasks separately.
To achieve real-time MRE down to single-shot spiral acquisition, we trained our model on undersampled data with 1-3 spiral arms, demonstrating its feasibility with up to 19-fold acceleration in data acquisition. This significant decrease in acquisition time minimizes the effects of motion and enables high frame-rate stiffness mapping, making it possible to track dynamic physiological processes. Furthermore, this approach facilitates free-breathing acquisitions, which are particularly beneficial for applications such as abdominal MRE or pediatric scans.

\subsection{In Vivo \& Different Number of Spiral Arms}
By varying the data-acquisition protocol using different acceleration factors, we demonstrated that, as anticipated, increasing the number of spirals resulted in progressively improved performance, as shown in Table \ref{tab:real_data_metrics}. Testing on in vivo data further validated the robustness of our model, despite inevitable variations in spiral trajectories and CSMs that lead to different undersampling artifacts than those in the simulated training data. Despite these differences, the proposed model demonstrated strong generalizability and maintained high performance (Fig. \ref{fig:SWS_real}), confirming its feasibility for potential clinical and real-world applications.

\subsection{Limitations}
While our proposed method demonstrates promising results, several limitations must be considered. First, the training process for our deep learning model requires significant computational resources and time, which can limit its practical application, particularly in resource-constrained environments such as clinical settings. Additionally, like all supervised training-based methods, the approach heavily depends on the availability of high-quality training data, making it difficult to apply in new contexts where such data are scarce or difficult to acquire. As a result, transferring the method to other body parts may present a significant challenge. However, as we have shown, it is possible to train on simulated data and successfully apply the model to newly acquired real data, potentially alleviating some of these limitations. An additional possible solution to this problem could be given by self-supervised approaches that do not require fully sampled reference data, as demonstrated in \cite{yaman2022zeroshot,martinssdu}. 
Another limitation of this study is the potential presence of measurement errors already present in the reference phase images acquired with 25 spirals, which could likely originate from inaccuracies in the acquisition process, given that spiral trajectories are particularly sensitive to gradient imperfections. These errors may affect the reference SWS maps seen in Fig. \ref{fig:SWS_real}. Addressing these errors, for example, through gradient impulse response function (GIRF) correction, is beyond the scope of this work but will be pursued in future studies.
A further limitation is that the evaluation was conducted exclusively on healthy volunteer data. Nevertheless, owing to the data-consistent design of the reconstruction framework and the use of an established stiffness estimation method, we consider it unlikely that the approach would fail in patient cohorts with abnormal anatomy or pathological stiffness. A systematic clinical validation in a larger patient population will, however, be an essential next step to confirm its broader applicability.

\subsection{Future Work} 
Future work could explore extending our approach to 3D, which would be computationally feasible with $T=1$ and particularly relevant for addressing the more complex wave fields observed in brain elastography. Further, an exciting direction would be integrating learnable components within the $k$-MDEV framework -- for instance, replacing fixed filters with a trainable counterpart or incorporating a data-driven strategy for phase-unwrapping. Additionally, regularization during fine-tuning could be improved -- for example, by adding a term penalizing the difference between the weights in pre-trained and fine-tuned models. Alternatively, a LoRA-like approach\cite{hu2022lora} could be employed, where a low-rank trainable matrix is added parallel to the frozen pre-trained model instead of fine-tuning all parameters, see \cite{yismaw2024domain} for an application to MRI.
Moreover, our model utilizes the temporal dimension as the primary source of redundant information for regularization. Expanding on this, future work could investigate leveraging correlations within the MEG and frequency dimensions to refine the regularization framework. Further, while in this work we followed an algorithm unrolling strategy, based on the observation that training times amounted to several days, one could opt for other approaches such as Plug and Play \cite{venkatakrishnan2013plug} or Regularization by Denoising \cite{romano2017little}, which employ off-the-shelf Gaussian denoisers based on deep NNs, see e.g.\ \cite{dong2018denoising}, \cite{kamilov2023plug}. 
Lastly, while this work focused on rapid brain MRE, future research could apply similar techniques in the abdomen, e.g., the liver, where managing respiratory motion artifacts is critical for the detection and staging of liver fibrosis\cite{kennedy2018quantitative}.

%% file: sections/7_conclusion.tex
\section{Conclusion} \label{sec:conclusion}
We introduced a novel method for estimating SWS maps from highly undersampled $k$-space data by integrating the $k$-MDEV inversion method as a trainable module within a CNN-regularized image reconstruction framework. This data-driven regularization approach demonstrated superior performance in both qualitative and quantitative image reconstruction, outperforming conventional methods such as TV regularization, as shown on a retrospectively undersampled brain dataset. Crucially, embedding $k$-MDEV into the model enables end-to-end optimization of network parameters for the ultimate clinical goal: accurate estimation of tissue mechanical properties via SWS mapping, rather than simply reconstructing intermediate image representations. Although trained exclusively on retrospectively undersampled data, our model successfully generalized to in vivo applications, underscoring its potential for clinical deployment. In particular, it enabled a 19-fold acceleration in data acquisition, facilitating real-time MRE and broader clinical adoption. Future work will explore self-supervised learning strategies to alleviate the dependence on large, labeled training datasets and enhance scalability.